\documentclass[10pt,journal,compsoc]{IEEEtran}

\usepackage[utf8]{inputenc}
\usepackage[english]{babel}
\usepackage{fullpage}

\usepackage{amsmath,amsfonts,amssymb}

\usepackage{array}

\usepackage[pdftex]{graphicx}
\usepackage{fixltx2e}
\usepackage{stfloats}

\ifCLASSOPTIONcompsoc
  \usepackage[nocompress]{cite}
\else
  \usepackage{cite}
\fi

\usepackage{xcolor}
\newcommand{\comment}[1]{\textcolor{red}{#1}}
\newcommand{\review}[1]{\textcolor{blue}{#1}}

\begin{document}
\title{Control Distance and Energy Scaling of Complex Networks}
\author{Isaac~Klickstein, Francesco~Sorrentino%
\IEEEcompsocitemizethanks{\IEEEcompsocthanksitem I. Klickstein and F. Sorrentino are with the Department of Mechanical Engineering at the University of New Mexico\protect\\
E-mail: iklick@unm.edu}
}
%
%
\IEEEtitleabstractindextext{%
\begin{abstract}
It has recently been shown that the average energy required to control a subset of nodes in a complex network scales exponentially with the cardinality of the subset.
While the mean scales exponentially, the variance of the control energy over different subsets of nodes is large and has as of yet not been explained.
Here, we provide an explanation of the large variance as a result of both the length of the path that connects control inputs to the target nodes and the redundancy of paths of shortest length.
Our first result provides an exact upper bound of the control energy as a function of path length between driver node and target node along an infinite path graph.
We also show that the energy estimation is still very accurate even when finite size effects are taken into account.
Our second result refines the upper bound that takes into account not only the length of the path, but also the redundancy of paths.
We show that it improves the upper bound approximation by an order of magnitude or more.
Finally, we lay out the foundations for a more accurate estimation of the control energy for the multi-target and multi-driver problem.
\end{abstract}

}

\maketitle

\IEEEdisplaynontitleabstractindextext
\IEEEpeerreviewmaketitle
%
%
\IEEEraisesectionheading{\section{Introduction}\label{sec:introduction}}
\IEEEPARstart{M}{uch} research has been performed in recent years concerning the control of complex networks \cite{liu2011controllability, liu2016control, ruths2014control, jia2013emergence, gao2016control, gao2014target, yuan2013exact, iudice2015structural}.
Controlling complex networks has numerous applications from regulating power grids \cite{pagani2013power}, routing traffic on the internet \cite{yan2006efficient}, marketing on social media \cite{proskurnikov2016opinion}, synchronization of multi-agent systems \cite{olfati2007consensus}, and many others.
Particular focus has been placed on the minimum control energy, and how it is affected by structural and dynamical properties of the network \cite{yan2015spectrum,yan2012controlling,klickstein2017energy}.
Due to multiple definitions in the field, we specify here that driver nodes refer to those nodes in the network which directly receive a control input signal and that target nodes refer to those nodes whose final state is prescribed by the control problem \cite{liu2016control, klickstein2017energy}.
Nodes may be either a driver node, a target node, both a driver and a target, or neither.\\
%
%
%
\indent 
Previous work has described how the control energy scales exponentially with the number of drivers \cite{yan2012controlling, yan2015spectrum} or the number of targets of the control action \cite{klickstein2017energy, shirin2017optimal, gao2014target}.
While these studies are based on a mix of numerical and theoretical arguments \cite{klickstein2017energy, yan2015spectrum}, very few analytical results have been obtained on the minimum control energy for a complex network.
Moreover, the exponential scaling is typically seen only after averaging over many realization of different distributions of control inputs or target nodes over the network.
While the mean of the control energy can be predicted, the variance is almost always quite large making predictions of the control energy for a particular distribution of target nodes and driver nodes over the network rather difficult.
To explain the large variance, we derive the analytical expression of the controllability Gramian for an infinite path graph.
We found that one cause of the large variance is that the required minimum control energy scales exponentially with the distance, defined as the length of the shortest path, between a driver node and a target node.
While the distance between driver nodes and target nodes has been known to play an important role for the control energy \cite{wang2012optimizing,chen2016energy}, we are able to provide the analytical reason behind the scaling.
Thus, when one chooses a random set of driver nodes and a random set of target nodes, the resulting control energy is dominated by the few target nodes at the greatest distance from any driver node, the number of which can vary greatly across multiple realizations.\\
\indent
A major difficulty in describing the control energy scaling more precisely is the fact that the network topology appears in the expression for the control energy.
Here, we present a methodology which separates the problem into two components, an expression for control energy independent of the network topology and topological properties of the graph's nodes.
In order to investigate more precisely the scaling of control energy of a particular realization, in this paper, we focus primarily on the single input and single target control case on a network.
We derive an exact solution for an  upper bound of the control energy for the single input, single target problem in a large network.
We also derive a relation which more accurately predicts the control energy that takes into account not only the distance, but also the path redundancy between input and target.
Finally, we discuss applications to the multi-driver, multi-target problem and describe how our results can be applied to general problems regarding the controllability of complex networks.\\
\section{Background}
\subsection{Control of Complex Networks}
Throughout this work, we consider undirected, unweighted graphs, unless otherwise specified, with associated adjacency matrix, $A$, where if $a_{ij} = 1$ ($a_{ij} = 0$), there exists an edge between nodes $i$ and $j$ (there does not exist an edge between nodes $i$ and $j$).
Also, we consider a uniform regulation parameter, $a_{ii}=-p$, that appears on the diagonal of the adjacency matrix that could possibly be equal to zero.
While the requirement that the network be undirected and unweighted is quite restrictive, we will preview how this work can be extended to both directed and weighted graphs in the discussion section.
We consider the dynamical equations that govern the evolution of the network as a system of linear time-invariant ODEs,
\begin{equation}\label{eq:linsys}
  \begin{aligned}
    \dot{\textbf{x}}(t) &= A \textbf{x}(t) + B \textbf{u}(t)\\
	\textbf{y}(t) &= C \textbf{x}(t)
  \end{aligned}
\end{equation}
where $\textbf{x}(t)$ is the vector of length $n$ of time-varying states for each node, $\textbf{u}(t)$ is the vector of length $n_d$ of control inputs, $\textbf{y}(t)$ is the vector of length $n_t$ of outputs, or targets as we call them, and $B$ and $C$ are the $n \times n_d$ input and $n_t \times n$ output matrices, respectively.
If the triplet $A,B,C$ is output controllable, then one can define the minimum control energy input that drives the network from an initial condition $\textbf{x}(t_0) = \textbf{x}_0$ to a prescribed final output $\textbf{y}(t_f) = \textbf{y}_f$ \cite{klickstein2017energy},
\begin{equation}
  \begin{aligned}
  \textbf{u}(t) &= B^T e^{A^T(t-t_f)} C^T \left(CW(t_f)C^T\right)^{-1}\\
  &\times\left(\textbf{y}_f - C e^{A(t_f-t_0)} \textbf{x}_0\right) 
  \end{aligned}
\end{equation}
The control energy can be shown to equal the quadratic form,
\begin{equation}\label{eq:energy}
  E = \boldsymbol{\beta}^T \left(CW(t_f)C^T\right)^{-1} \boldsymbol{\beta}
\end{equation}
where $\boldsymbol{\beta} = \left( \textbf{y}_f - C e^{A(t_f-t_0)} \textbf{x}_0 \right)$ is defined as the control action, i.e., the difference between the prescribed output and the zero-input output at time $t_f$.
Also, the matrix $W(t_f)$ is the controllability Gramian which satisfies the linear Lyapunov differential equation,
\begin{equation}\label{eq:gram}
  \begin{aligned}
    \dot{W}(t) = AW(t) + W(t)A + BB^T, && W(0) = O_n
  \end{aligned}
\end{equation}
where $O_n$ is the matrix of all zeros of dimension $n \times n$, $A = A^T$ as the graph is undirected, and one can find $W(t_f)$ by integrating Eq. \eqref{eq:gram} forward until $t = t_f$.
If the matrix $A$ is Hurwitz, then Eq. \eqref{eq:gram} has a single stable fixed point which can be determined by solving the algebraic Lyapunov equation $O = AW + WA + BB^T$.
The steady state Gramian, $W$, is a function of only the adjacency matrix $A$ and the distribution of control inputs as described by the matrix $B$.
In the following analysis, we will assume that $W$ is the fixed point Gramian which represents the control energy in the large time limit and $W(t)$ is the finite time controllability Gramian.\\
\indent
Note that the control energy in Eq. \eqref{eq:energy} is a function of \emph{both} the topology of the network as represented by $A$ (which appears in the expression for $W(t_f)$) and the distribution of inputs and outputs as represented by $B$ and $C$, respectively.
In the analysis that follows, we assume that there are $n_d$ driver nodes ($n_t$ target nodes) so that there are $n_d$ columns in $B$ ($n_t$ rows in $C$) where each column (row) is a versor with the single non-zero element corresponding to the index of a driver (target) node.
With this definition of $C$, the output Gramian $CW(t_f)C^T$ is a principal submatrix of $W(t_f)$ corresponding to the target node indices.
Determining the control energy requires (1) computing the $n \times n$ controllability Gramian and (2) performing the inversion of the output Gramian $CW(t_f)C^T$.
In the following sections we present a methodology that allows one to instead approximate the energy using (1) structural properties of the graph which need only be found once (as opposed to the need to recalculate $W(t_f)$ for different choices of the matrix $B$) and (2) energy properties of a much smaller graph (up to about 20 nodes) which also needs to only be calculated once (as opposed to the need to re-invert the output Gramian each time one changes the particular choice of target nodes). 
%
%
\subsection{Single Input Single Output Problem}
\indent
For the single input problem, the matrix $B$ is a vector with a single non-zero element equal to one.
If the single non-zero element is located at position $B_k$, then the control input is connected to node $k$ making node $k$ the sole driver node.
Similarly, for the single target problem, the matrix $C$ is a vector with only a single non-zero element equal to one.
If the single non-zero element is located at position $C_{\ell}$, then node $\ell$ is the sole target of the control action.
The control energy for the single target problem can be simplified to,
\begin{equation}\label{eq:longtime}
  E = \beta_{\ell}^2 \frac{1}{W_{\ell,\ell}(t_f)}
\end{equation}
where $\beta_{\ell}$ is the $\ell$th component in the vector $\boldsymbol{\beta}$ as defined previously and $W_{\ell,\ell}(t_f)$ is the $\ell$th diagonal component in the matrix $W(t_f)$ as defined in Eq. \eqref{eq:gram}.
%
%
\section{Infinite Path Graph}
Previously \cite{klickstein2017energy,yan2015spectrum} it has been shown that the variance of the control energy over different sets of driver nodes and target nodes of constant sizes $n_d$ and $n_t$ is large.
We begin analyzing the cause of the large variance by setting $n_d = n_t = 1$ and finding an analytical expression for an upper limit of the control energy as a function of the structural properties of the graph.
We also hypothesize that the large variation in control energy for any particular distance (length of the shortest path) between a driver node and a target node is due to multiple (redundant) paths between the driver node and the target node and that the presence of these redundant paths reduces the control energy.
For each control distance then, the expected maximum control energy for a particular distance between driver node and target node corresponds to a graph with only a single path between them, which we call a path graph.\\
%
%
\begin{figure}[t!]
  \centering
  \includegraphics[width=\columnwidth]{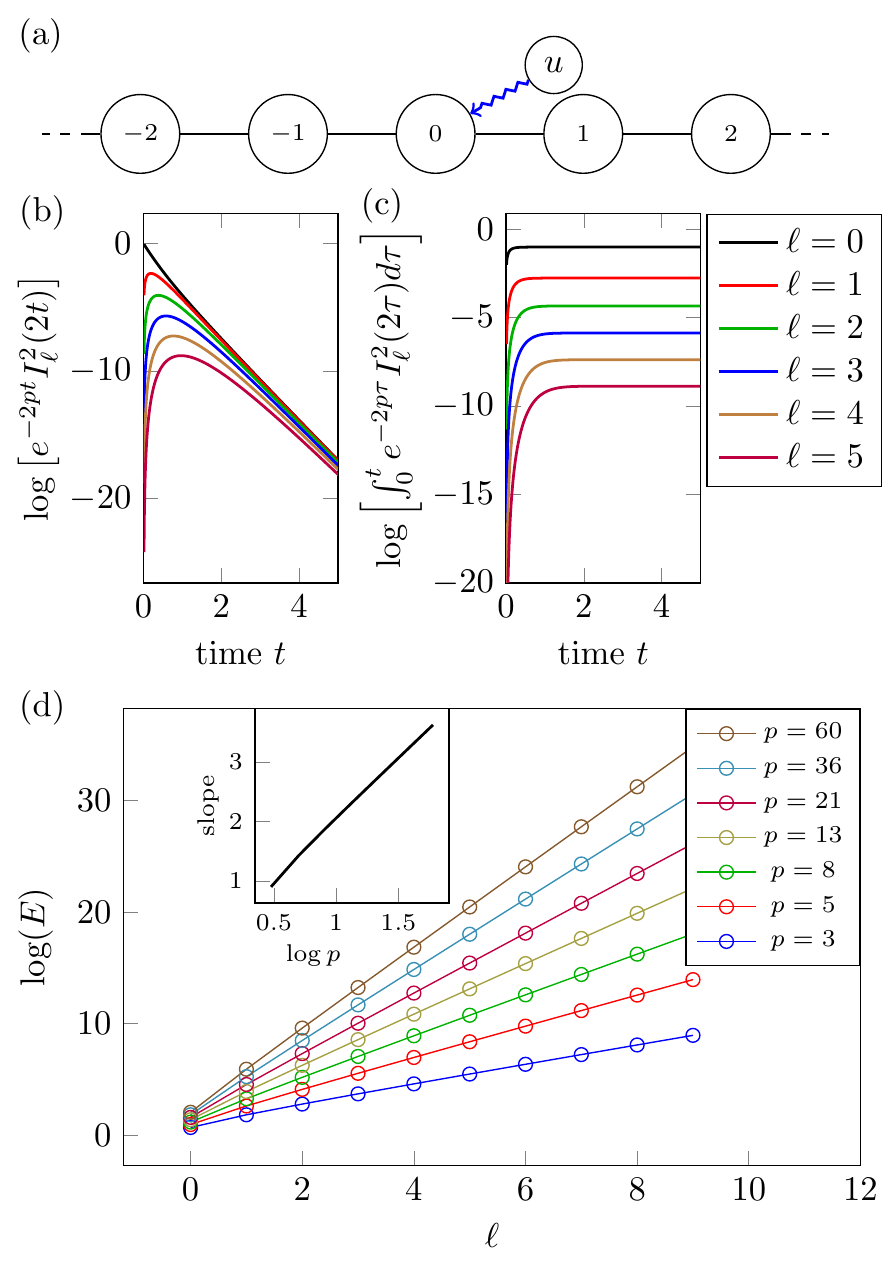}
  \caption{
    Details of the path graph in terms of energy.
    (a) a diagram of the infinite path graph with the single control input attached to the node labeled $0$ and additional nodes emanating from either side.
    Typical time evolutions of the integrand (b) and integral (c) in Eq. \eqref{eq:chaingram} ($p=7$, $f=1$) for different distances between driver node and target node, $\ell$.
    (b) \review{As time grows} larger, the exponential term dominates and we see an exponential decay approximately equal to $e^{-2pt}$.
    (c) with the exponential decay of the integrand, the integrals converge to a constant value which is expected as the adjacency matrix when $p=7$ and $f=1$ is Hurwitz.
    (d) energy curves for path graphs with different regulation parameters $p$.
    We see that for any value $p$, the energy increases exponentially with distance between input and target, $\ell$.
    As $p$ increases, the rate of increase (slope) of energy increases as well.
    This is shown in the inset where we see the slope of the energy curve increases exponentially with the log of the magnitude of the self-regulation.
    The rate of increase of the slopes is approximately $2$.
  }
  \label{fig:path}
\end{figure}
%
%
\begin{figure}[t!]
  \centering
  \includegraphics[width=\columnwidth]{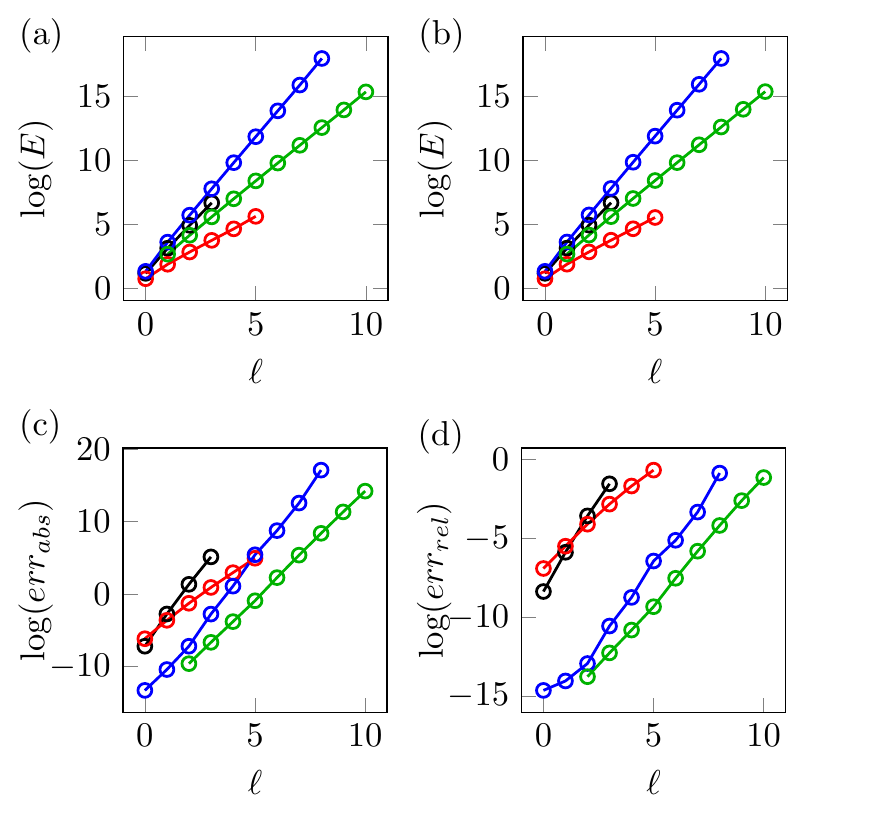}
  \caption{
    Finite effects on the energy of the path graph.
    Four finite path graphs are considered with properties: $\ell_{\max} = 3$ and $p=7$ (black), $\ell_{\max}=5$ and $p=3$ (red), $\ell_{\max} = 8$ and $p = 10$ (blue), and $\ell_{\max} = 10$ and $p=5$ (green).
    The infinite energy curves (a) and the finite energy curves (b) appear to be qualitatively similar.
    The absolute error (c) only grows large ($\geq 1$) when the target is closer to the end of the finite chain than to the driver node.
    However, the relative error (d) though remains small until the target node is at the very end of the chain.
    Overall, the infinite chain adequately predicts the order of magnitude of the energy (small relative error) but may do poorly when trying to predict the precise value (when the absolute error becomes large).
  }
  \label{fig:finite}
\end{figure}
\begin{figure}[t!]
  \centering
  \includegraphics[width=\columnwidth]{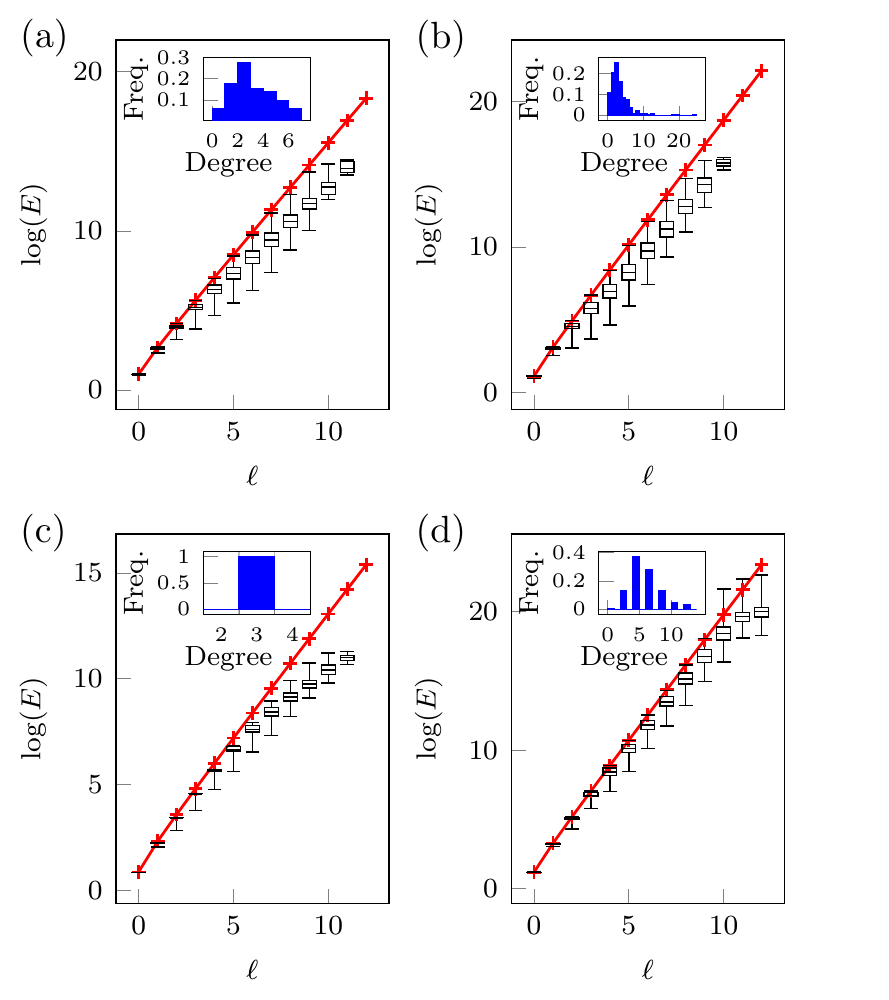}
  \caption{
    Comparison of control energy between the infinite path graph and finite graphs of various types.
    The graphs considered are (a) an ER graph of $n=300$ and $p=0.01$ (b) a scale-free graph of $n=300$ and $\gamma = -3$ (c) a $3$-regular graph of size $n=300$, and (d) the Northern European power grid.
    Each graph has regulation parameter $p$ chosen such that the maximum eigenvalue of the adjacency matrix is $\lambda_{\max} = -1$.
    All insets show the degree distribution of the particular graph considered.
    The red curves are the control energy of an infinite path graph with regulation parameter equal to that of the particular graph.
    The boxplots show the median, lower and upper quantiles, and minimum and maximum control energy for each distance $\ell$ between control input and target node.
    Note that for most cases, the maximum control energy for each distance is quite close to the corresponding control energy of the infinite graph. 
    Also, note that the majority of control energy values for each distance are considerably less than the predicted value by the infinite path.
  }
  \label{fig:path_compare}
\end{figure}
\begin{figure}[t!]
  \centering
  \includegraphics[width=\columnwidth]{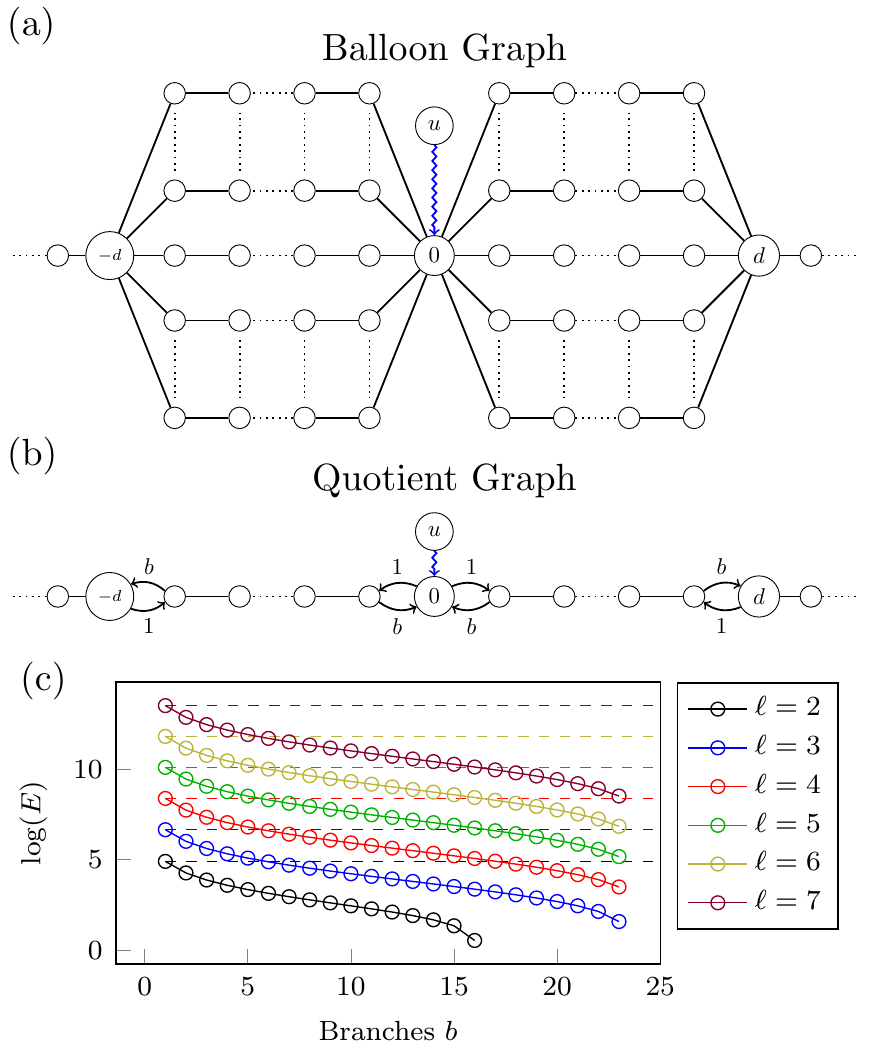}
  \caption{
    Demonstration of properties of the balloon graph.
    a) a diagram of the balloon graph.
    The central node is labeled $0$, and receives the single control input.
    Emanating from node $0$ are $b$ branches in each direction.
    Each branch is of length $d-1$ before recombining into the nodes labeled $d$ and $-d$, where we take node $d$ to be the target node.
    b) the associated quotient graph which is a bidirectional path graph with four non-unary edge weights located at $A_{0,1}$, $A_{0,-1}$, $A_{d,d-1}$, and $A_{-d,-d+1}$, all of value $b$.
    c) the control energy for various values of distance $\ell$ (the colored curves) and number of branches (along the horizontal axis).
    All balloon graphs are constructed with self-regulation parameter $p = -7$.
    As the number of branches increases the control energy decreases, but not uniformly.
    Dashed lines correspond to the infinite path ($b=1$) of the same regulation parameter $p$ and same control distance $\ell$.
    The curves terminate when the number of branches becomes large enough such that the adjacency matrix is no longer Hurwitz.
  }
  \label{fig:balloon}
\end{figure}
\indent
Consider an infinite, bidirectional path graph, as represented in Fig. \ref{fig:path}(a), so that the nodes are numbered from $-\infty$ to $\infty$ with node $0$ acting as the single driver node and the single target node, $\ell$, lying somewhere along the path.
Every edge in the graph between nodes $i$ and $i+1$ (or vice versa) has weight $f$.
Each node also has a self-loop with value $-p$, which appears as the uniform diagonal values in the adjacency matrix $A$.
The evolution of an element of the controllability Gramian for the path graph can be simplified to,
\begin{equation}\label{eq:diffchain}
  \begin{aligned}
    \dot{w}_{i,j}(t) &= -2p w_{i,j}(t) + fw_{i-1,j}(t) + fw_{i+1,j}(t)\\
    &+ fw_{i,j-1}(t) + fw_{i,j+1}(t) + \delta_{i,0} \delta_{j,0},\\ -\infty &< i,j < \infty
  \end{aligned}
\end{equation} 
where $\delta_{i,j}$ is the Kronecker delta and the initial condition $w_{i,j}(0) = 0$ for all $i$ and $j$.
To derive the solution to Eq. \eqref{eq:diffchain}, we first solve the homogeneous problem, i.e., we neglect the $\delta_{i,0}\delta_{j,0}$ term, and then second we include the driver node portion.
To solve for the time evolution of the homogeneous controllability Gramian of the infinite path graph, we apply a two-variable discrete-time Fourier transform to Eq. \eqref{eq:diffchain} (neglecting $\delta_{i,0}\delta_{j,0}$),
\begin{equation}
  w_{u,v}(t) = \sum_{i,j} e^{\mathcal{I}ui} e^{\mathcal{I}vj} w_{i,j}(t)
\end{equation}
where $\mathcal{I} = \sqrt{-1}$.
This transformation allows us to decouple the Gramian equation so that,
\begin{equation}\label{eq:Wuv}
  \dot{w}_{u,v}(t) = (-2p + 2f\cos u + 2f \cos v) w_{u,v}(t)
\end{equation}
As Eq. \eqref{eq:Wuv} is simply a scalar, linear, ODE, its solution is,
\begin{equation}
  w_{u,v}(t) = e^{-2pt} e^{2ft \cos u} e^{2ft \cos v} w_{u,v}(0)
\end{equation}
where the initial condition in $u,v$ space can be written in terms of $i,j$ space giving the final form,
\begin{equation}
  \begin{aligned}
  w_{u,v}(t) &= e^{-2pt} e^{2ft \cos u} e^{2ft \cos v} \sum_{\alpha,\beta} e^{\mathcal{I}u\alpha} e^{\mathcal{I}v\beta} W_{\alpha,\beta}(0)
  \end{aligned}
\end{equation}
Performing the inverse DTFT transform to return to $i,j$ space,
\begin{equation}\label{eq:homogeneous}
  \begin{aligned}
    w_{i,j}(t) &= e^{-2pt} \sum_{\alpha,\beta} \frac{1}{4\pi^2} \int_{-\pi}^{\pi} \int_{-\pi}^{\pi} e^{-\mathcal{I}u(i-\alpha)} e^{-\mathcal{I}v(j-\beta)}\\
    &\times e^{2ft \cos u} e^{2ft \cos v} du dv\\
    &= \sum_{\alpha,\beta} e^{-2pt} I_{i-\alpha}(2ft) I_{j-\beta}(2ft) w_{\alpha,\beta}(0)
  \end{aligned}
\end{equation}
where $I_{\ell}(z)$ is the Modified Bessel Function of the First Kind (MBFFK) of integer order.
\begin{equation}
  I_{\ell}(z) = \frac{1}{2\pi} \int_0^{\pi} e^{z\cos\theta}\cos(\ell \theta) d\theta
\end{equation}
Returning to the Gramian expression in Eq. \eqref{eq:diffchain}, we can include an additional term that represents the locations of, possiby multiple, driver nodes defined as the set $\mathcal{D}$ of integers.
\begin{equation}
  g = \sum_{k \in \mathcal{D}}\delta_{i,k}\delta_{j,k}
\end{equation}
As the Gramian equation is linear, let $\eta_{i,j}(t)$ represent the homogeneous solution in Eq. \eqref{eq:homogeneous},
\begin{equation}
  \eta_{\substack{i-\alpha\\j-\beta}}(t) = e^{-2pt}I_{i-\alpha}(2ft)I_{j-\beta}(2ft)
\end{equation}
so the non-homogeneous Gramian elements are (noting $w_{i,j}(0) = 0$),
\begin{equation}
  \begin{aligned}
  w_{i,j}(t) &= \sum_{\alpha,\beta} \eta_{\substack{i-\alpha\\j-\beta}}(t) W_{\alpha,\beta}(0)\\
  &+ \int_0^t \sum_{\alpha,\beta}\eta_{\substack{i-\alpha\\j-\beta}}(\tau) g d\tau\\
  &= \int_0^t \sum_{k \in \mathcal{D}} \eta_{\substack{i-k\\j-k}}(\tau) d\tau
  \end{aligned}
\end{equation}
Finally, for the particular case when there is one input at node $0$, i.e., $\mathcal{D} = \{0\}$ and $g = \delta_{i,0}\delta_{j,0}$ as is the case in Eq. \eqref{eq:diffchain}, we get the form used for the remaining analysis of the infinite path graph,
\begin{equation}\label{eq:chaingram}
  \begin{aligned}
    &w_{i,j}(t) = \int_0^t e^{-2p\tau} I_{i}(2f\tau) I_j(2f\tau) d\tau, \\ &-\infty < i,j < \infty
  \end{aligned}
\end{equation}
For the single target problem, we are concerned with only the diagonal element,
\begin{equation}\label{eq:Wll}
  w_{\ell,\ell}(t) = \int_0^t e^{-2p\tau} I_{\ell}^2(2f\tau) d\tau.
\end{equation}
To compare the results of Eq. \eqref{eq:Wll} to real networks, we assume that the regulation parameter $p$ is large enough in magnitude so as to make the adjacency matrix $A$ Hurwitz.
While Eq. \eqref{eq:longtime} provides the exact value for the single target energy, we can approximate an upper bound for the large time energy for a particular control distance $\ell$ as,
\begin{equation}\label{eq:diag}
  E_{\ell}^{UB} \simeq \beta_{\ell}^2 \left[ \int_0^{\infty} e^{-2p\tau} I_{\ell}^2 (2f\tau) d\tau \right]^{-1}
\end{equation}
where the approximation occurs because the expression in Eq. \eqref{eq:Wll} is only exact in the infinite path graph case.
We present some typical curves of the integrand and integral in Eq. \eqref{eq:diag} in Figs. \ref{fig:path}(b) and \ref{fig:path}(c).
As the MBFFK is an increasing function of its argument $\tau$, while the exponential term is decreasing, we see a peak in the curves in Fig. \ref{fig:path}(b).
The integral in Eq. \eqref{eq:diag} decreases as the distance $\ell$ increases, as shown by the steady state behavior in Fig. \ref{fig:path}(c).
This implies that the energy increases with increasing distance between driver node and target node as we expected.
The control energy is further explored in Fig. \ref{fig:path}(d), where we see, for increasing values of $p$, the rate of increase of the energy with respect to distance $\ell$ also increases.
The energy can be thought of as a function of the distance between driver node and target node $\ell$ and the self-regulation parameter $p$, $E(\ell,p)$.
Then, by defining $s(p)$ as the slope of the energy curves in Fig. \ref{fig:path}(d), we can compute,
\begin{equation}
  \begin{aligned}
    s(p) &= \frac{\partial}{\partial \ell} \log E(\ell,p)\\
    &= -\frac{2}{\ln 10}\frac{\int_0^{\infty} e^{-2pt} I_{\ell}(2ft) \frac{\partial}{\partial \ell} I_{\ell}(2ft) dt}{\int_0^{\infty} e^{-2pt} I_{\ell}^2 (2ft) dt}
  \end{aligned}
\end{equation}
From the inset in Fig. \ref{fig:path}(d), we also see that $\frac{\partial s(p)}{\partial \log_{10} (p)} \approx 2$.\\
\indent
While the energy curves of the path graphs presented are exact in the limit of the infinite path, we now examine how finite size effects reduce the accuracy of Eq. \eqref{eq:chaingram}.
The energy curves for four finite path graphs of varying lengths, $\ell_{\max}$, and regulation parameter $p$ are shown in Fig. \ref{fig:finite}(a) computed using Eq. \eqref{eq:longtime}.
Comparable energy curves (for same range of $\ell$ and value $p$) for infinite path graphs are shown in Fig. \ref{fig:finite}(b) computed using Eq. \eqref{eq:diag}.
The absolute errors between the finite and infinite path graph energy curves are shown in Fig. \ref{fig:finite}(c).
The relative error, defined as the absolute error divided by the control energy of the finite path graph, remains small ($<1$) as shown in Fig. \ref{fig:finite}(d) which means that the infinite path graph is able to predict the order of magnitude of the energy of the finite path.\\
\indent
We now turn to other finite graphs with complex topology.
In Fig. \ref{fig:path_compare}, we compare the single driver node single target node energy curves from the infinite path graph to single driver node single target node energy values in a selection of finite graphs.
The red curves in each panel are the infinite path energy curves computed using Eq. \eqref{eq:diag}.
The box and whisker marks represent the distribution of energy values for all pairs of nodes in a particular network for each distance between driver and target nodes.
The lower and upper bounds of each box correspond to the 25\% and 75\% quantiles of the data, respectively, and the lower and upper bars correspond to the minimum and maximum values, respectively.
The mark near the middle of each box is the median value.
For the Erdos Renyi (Fig. \ref{fig:path_compare}(a)) and scale free (Fig. \ref{fig:path_compare}(b)) networks, the infinite path provides a good approximation of the maximum energy values.
For the $k$-regular (Fig. \ref{fig:path_compare}(c)) network, we typically see that the infinite path graph energy substantially over-estimates the control energy when the distance between driver node and target node grows large.
Finally, as an example of a real network, the single driver single target simulation is performed on the Northern European power grid \cite{menck2014dead} (Fig. \ref{fig:path_compare}(d)) and the results are again compared to the infinite path graph.
The estimated control energy provided by the infinite path both over-estimates and very rarely under-estimates  the maximum value for different control distances $\ell$.
It has been shown that the Northern European power grid has a number of `dead-ends' \cite{menck2014dead} which are essentially finite path graphs emanating from a denser part of the graph.
Thus, for at least one pair of nodes, we recreate the worst case scenario (targeting a node at the end of a finite path) as included in the discussion of Fig. \ref{fig:finite}.
For all four cases though, the vast majority of the energy values are around an order of magnitude less than the infinite path graph.
Overall, for large networks, we see that Eq. \eqref{eq:chaingram} provides a good approximation of the maximum expected energy for a particular distance between driver node and target node independent of any network topology.\\
\section{Redundant Paths}
\indent
As previously mentioned, we hypothesize that often the energy of the single driver single target problem is less than the predicted maximum of the path graph due to the redundancy of shortest paths between driver node and target node.
To understand how the redundancy of paths affects the control energy, we define the \emph{balloon graph} as a modification to the path graph.
A balloon graph can be completely defined by two integer values, $b$ which is the number of redundant paths and $d$ which is the length of the redundant paths before they converge back to nodes $d$ and $-d$ as shown in Fig. \ref{fig:balloon}(a).
We set node $0$ to be the single driver node and node $d$ to be the single target node, labeled in Fig. \ref{fig:balloon}(a).
Before examining the improvements on predicting the energy value, we must define graph symmetries and quotient graphs.\\
\begin{figure}[t!]
  \centering
  \includegraphics[width=\columnwidth]{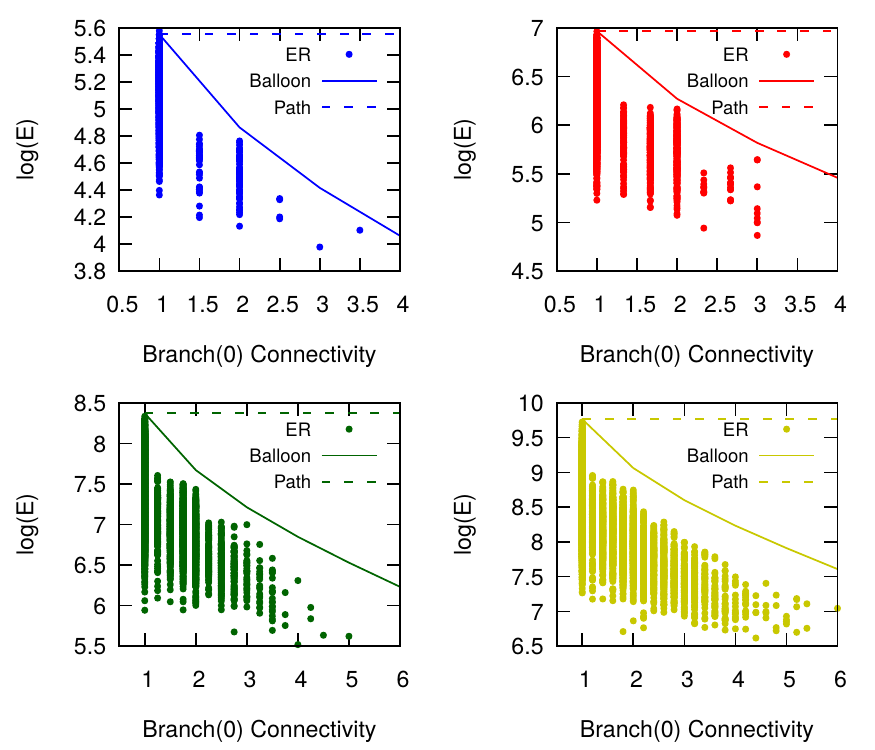}
  \caption{
    Comparison between the infinite path graph, balloon graph, and an ER graph.
    For control distances (a) $\ell = 3$, (b) $\ell = 4$, (c) $\ell = 5$, and (d) $\ell = 6$, we see the path graph (dashed line) greatly over-estimates the control energy for most of the values.
    On the other hand, the control energy is much better approximated by the balloon graph (solid lines) which captures the decrease as the redundancy of paths increases.
    Despite this, most values are still considerably less, which suggests that redundancy of paths longer than the shortest path must also play an important role.
  }
  \label{fig:balloon_compare}
\end{figure}
\indent
Let $\mathcal{G} = (\mathcal{V},\mathcal{E})$ be a graph with $|\mathcal{V}|=n$ nodes and $|\mathcal{E}|=m$ edges.
The symmetry group associated with the graph $\mathcal{G}$ is the set of permutations that act on the node set that preserve the set of edges, that is,
$\mathcal{G}' = (\pi(\mathcal{V}),\mathcal{E}) = \mathcal{G}$.
One can also define the symmetry group associated with a node colored graph such that no two nodes may be permuted if they are given different colors.
The symmetry group partitions the set of nodes into $n_o$ disjoint subsets, called orbits, where each orbit consists of nodes that may be permuted among each other.
We assume that orbit $i$ consists of $n_i$ nodes.
From now on, we will assume that the nodes are labeled such that the first $n_d$ nodes are driver nodes, each in their own orbit, and all other nodes are labeled such that nodes in the same orbit have adjacent labels.
This labelling allows us to write the adjacency matrix in block orbit form such that block $A_{ij}$ describes how the nodes in orbit $i$ receive edges from the nodes in orbit $j$.
Every node in orbit $i$ receives the same number of incoming edges from the nodes in orbit $j$ for all orbits ( see e.g. \cite{pecora2014cluster}).
This leads to the important property that the row-sums of orbit block $A_{ij}$ are all equal,
\begin{equation}
  A_{ij} \boldsymbol{1}_{n_j} = \bar{a}_{ij} \boldsymbol{1}_{n_i}
\end{equation}
where $\boldsymbol{1}_n$ is the vector of all ones of length $n$.
\indent
One can now define a quotient graph $\mathcal{Q}$ where each node represents an orbit in the original graph and the edges between nodes in the quotient graph have weights corresponding to the number of edges a node in orbit $i$ receives from nodes in orbit $j$ (i.e., the row-sum $\bar{a}_{ij}$).
The relationship between the adjacency matrix of the original graph $A$ and the adjacency matrix of the quotient graph $A_Q$ is,
\begin{equation}
  A^Q = H^+ A H
\end{equation}
where the $n \times n_o$ matrix $H$ is the indicator matrix with elements $H_{ik} = 1$ if node $i$ is in orbit $k$ and $H_{ik} = 0$ otherwise and $H^+ = (H^T H)^{-1} H^T$ is the Moore-Penrose pseudoinverse of the matrix $H$ \cite{schaub2016graph}.\\
\indent
A symmetric permutation, as described above, can be represented as a permutation matrix $P$ with the property $PAP^T = A$, $PB = B$, and $PP^T = I$.
Applying this type of permutation to Eq. \eqref{eq:gram} allows us to write the Gramian in terms of a permuted network such that $\tilde{W}(t) = PW(t)P^T$.
\begin{equation}
  \begin{aligned}
    P \dot{W}(t) P^T &= P W(t) P^TP A P^T\\
    &+ PAP^T P W(t) P^T + P BB^TP^T
  \end{aligned}
\end{equation}
or, equivalently,
\begin{equation}
  \begin{aligned}
    \dot{\tilde{W}}(t) &= \tilde{W} A + A \tilde{W}(t) + BB^T
  \end{aligned}
\end{equation}
We see that $\tilde{W}(t)$ satisfies the same equation as $W(t)$, in Eq. \eqref{eq:gram}, with the same initial condition, hence $\tilde{W}(t) = W(t)$, and, more specifically, if $i$ and $k$ are in the same orbit, and $j$ and $\ell$ are in the same orbit, then $w_{i,j}(t) = w_{k,\ell}(t)$ which implies the presence of redundant equations in Eq. \eqref{eq:gram}.
We can thus write the controllability Gramian in a similar block orbit form as we did for $A$ above so that the block $W_{ij}(t)$ is,
\begin{equation}
  W_{ij}(t) = w_{ij}(t) J_{n_in_j}
\end{equation}
where $w_{ij}(t)$ is single value of all elements in the block $J_{n_in_j}$ is the matrix of all ones of dimension $n_i \times n_j$, the populations of orbits $i$ and $j$.
It is possible to contract the Gramian matrix, and the differential equation describing its evolution in Eq. \eqref{eq:gram}, to just those unique values.
What remains is the contracted controllability Gramian, $V(t)$.
We now prove that $V(t)$ is the controllability Gramian associated with the quotient graph,
\begin{equation}
  \begin{aligned}
    (H^+ \dot{W}(t) H^{+^T})_{ij} &= H^+ AW H^{+^T} + H^+ WA H^{+^T}\\
    &+ H^+ BB^T H^{+^T}\\
    &= \frac{1}{n_in_j} \sum_{k=1} \bar{a}_{ik} w_{kj}(t) \boldsymbol{1}_{n_i}^T J_{n_in_j} \boldsymbol{1}_{n_j}\\
    &+ \frac{1}{n_in_j} \sum_{k=1} \bar{a}_{jk} w_{ik}(t) \boldsymbol{1}_{n_i}^T J_{n_in_j} \boldsymbol{1}_{n_j}\\
    &+ \frac{1}{n_in_j} b_{ij} \boldsymbol{1}_{n_i}^T J_{n_in_j} \boldsymbol{1}_{n_j}\\
    &= \sum_{k=1} \bar{a}_{ik} w_{jk}(t) + \sum_{k=1} w_{ik}(t) \bar{a}_{jk}\\
    &+ b_{ij}\\
    &= (A_Q V(t))_{ij} + (V(t) A_Q^T)_{ij}\\
    &+ (B_QB_Q^T)_{ij}\\
    &= \dot{V}_{ij}(t)
  \end{aligned}
\end{equation}
where $\dot{V}_{ij}(t)$ is a single scalar element of the controllability Gramian evolution equation for the quotient graph.\\
Thus, we can saw that,
\begin{equation}
  \dot{V}(t) = A_Q V(t) + V(t) A_Q^T + B_Q B_Q^T
\end{equation}
\indent
We use the set of vertex-colored symmetries to define the quotient graph of the balloon graph, shown in Fig. \ref{fig:balloon}(b), which is simply a path graph with four edge weights in the adjacency matrix at elements $A_{-d,-d+1}$, $A_{-1,0}$, $A_{0,1}$, $A_{d,d-1}$ each with weight $b$.
Note that the quotient graph of the balloon graph is simply a weighted infinite path graph and thus from the analysis in Fig. \ref{fig:finite}, we are able to approximate the controllability Gramian with a finite weighted path with only a few nodes.\\
\indent
Despite this very small change, the closed form solution that was available for the infinite path graph can no longer be determined.
However, we can still numerically compute the elements of the controllability Gramian for the quotient graph of the finite balloon graph, which is simply a weighted path graph, and which is computationally advantageous compared to more real and model networks of interest, which will have hundreds or thousands of nodes.
As will be seen, computing the controllability Gramian of this small weighted path graph provides two major benefits; (1) it is a more numerically efficient method to approximate the required control energy than computing the controllability Gramian of the large original network and (ii) it provides a more accurate approximation than the closed form solution of the infinite unweighted path graph.\\
\indent
With the above formulation, we compute the control energy for the single driver, single target problem with the balloon graph for varying distances $\ell$ and number of branches $b$ with sample regulation parameter $p = 7$.
In Fig. \ref{fig:balloon}(c) we see a rapid decrease at first for all values of $\ell$ before a period of stable decrease until the end of the curves which decrease rapidly again.
The curves terminate when the increasing number of branches leads to a non-Hurwitz adjacency matrix.
Dashed lines included represent the energy value for an equivalent path graph of length $\ell$ ($b=1$).
We see that when there are five or more redundant paths, the energy required for a node at distance $\ell$ is approximately equal to the energy of a path graph at distance $\ell-1$, which is also the type of behavior we have seen in large complex networks.\\
\indent
To compare the energy prediction of the energy path to large complex networks, we define the following metric.
Let $d_{i,j}$ be the length of the shortest path between nodes $i$ and $j$.
Let $\mathcal{V}_{i,j}$ be the set of nodes that lie on a shortest path between nodes $i$ and $j$, that is, if $k \in \mathcal{V}_{i,j}$ then $d_{ik} + d_{kj} = d_{ij}$.
We define the branch connectivity of two nodes $i$ and $j$ to be,
\begin{equation}
  B(i,j) = \frac{|\mathcal{V}_{i,j}|-1}{d_{ij}}
\end{equation}
Note that for a path graph $B(i,j) = 1$ and for a balloon graph with $b$ branches, $B(0,d) = b$.
For a complex network, $B(i,j)$ can take fractional values as well.
As a typical example, Fig. \ref{fig:balloon_compare} plots the single driver single target control energy for a path graph (dashed line), a balloon graph (solid line) and an Erdos-Renyi graph of size $n=300$ and edge probability $e_p = 0.02$ for different distances $\ell$ between driver node and target node.
We see that the balloon graph provides a tighter upper bound than the path graph sometimes by an order of magnitude or more.
For each value of branch connectivity we still see there is a large variation of values of control energy which suggests that redundant paths of length longer than the minimum distance must also play a role in a less obvious way.
\section{Multi-Driver and Multi-Target Problem}
More often we are interested in the case when there are multiple driver nodes.
The results of Eq. \eqref{eq:chaingram} for the infinite path graph can be extended to the case where multiple nodes along the path receive independent control signals \comment{REMOVED REFERENCE TO APPENDIX. IT STATED WE HAD THE DERIVATION OF EQ. \eqref{eq:multi_driver}}
Let $\mathcal{D}$ be the set of indices of the driver nodes so $|\mathcal{D}| = n_d$, then, the controllability Gramian has elements,
\begin{equation}\label{eq:multi_driver}
   w_{i,j}(t) = \int_0^t \sum_{k \in \mathcal{D}} e^{-2p\tau} I_{i-k}(2f\tau) I_{j-k}(2f\tau) d\tau
\end{equation}
For the single target problem, two immediate conclusions can be made from Eq. \eqref{eq:multi_driver}; (1) adding an additional driver anywhere along the infinite path always decreases the required energy (as $w_{i,j}(t)$ increases with the addition of driver nodes) and (2) driver nodes closer to the target node (i.e. smaller $|\ell-k|$) contribute much more to reducing the required energy than driver nodes further away.
This is in agreement with the recently published result \cite{summers2016submodularity} that certain metrics of the controllability Gramian are submodular with the addition of driver nodes.
They find that the trace, the negative trace of the inverse, and the logarithm of the determinant of the controllability Gramian are all submodular set functions with the addition of more driver nodes.
All of these metrics are either sums or products of the individual elements of the controllability Gramian.
On the other hand, they find that the smallest eigenvalue of the controllability Gramian is not submodular, which we also see here as simply increasing the individual elements of a matrix does not necessarily increase the eigenvalues of the matrix.\\
\indent
We now consider the general problem of multiple driver nodes and multiple \review{control inputs.}
Once again, consider the infinite path graph, now with multiple target nodes.
We can construct the output Gramian using either single driver elements in Eq. \eqref{eq:chaingram} or the multiple driver elements in Eq. \eqref{eq:multi_driver}.
Now, rather than the energy simply being found by examining the scalar values along the diagonal of the controllability Gramian as for the single target problem, we must examine the spectral properties of the output Gramian matrix.
Let $\mu_i$ be an eigenvalue of the output controllability Gramian $CWC^T$ and $\textbf{v}_i$ its corresponding eigenvector.
As $CWC^T$ is, by definition, a principle minor of the symmetric matrix $W$, it too is symmetric.
Sort the eigenvalues so that $\mu_1 \leq \mu_2 \leq \mu_3 \ldots$.
The control energy for some control action $\boldsymbol{\beta}$ can be written as,
\begin{equation}
  \begin{aligned}
    E &= \boldsymbol{\beta}^T \left(CWC^T\right)^{-1} \boldsymbol{\beta}\\
    &= \boldsymbol{\beta}^T \left(\sum_{i=1}^{n_t} \frac{1}{\mu_i} \textbf{v}_i \textbf{v}_i^T \right) \boldsymbol{\beta}\\
    &= \sum_{i=1}^{n_t} \frac{1}{\mu_i} \alpha_i^2
  \end{aligned}
\end{equation}
where $\alpha_i = \boldsymbol{\beta}^T \textbf{v}_i$ is the magnitude of the projection of the control action along the $i$th eigenvector of the output controllability Gramian.
It has been shown that typically, the smallest eigenvalue, $\mu_1$, of the output controllability Gramian is much smaller than $\mu_2$ \cite{klickstein2017energy}.
For an arbitrary control action $\boldsymbol{\beta}$, $\alpha_1 \neq 0$ and so typically the maximum term in the summation of the control energy is proportional to $\frac{1}{\mu_1}$.
We say that the worse-case energy occurs when the control action is parallel to the eigenvector associated with the smallest eigenvalue, $\mu_1$,
\begin{equation}
  E_{\max} = \frac{1}{\mu_1}
\end{equation}
When $n_t > 1$, one can generally characterize the control energy corresponding to the output controllability Gramian by its smallest eigenvalue $\mu_1$.\\
\indent
For concreteness, the case for two targets and one driver on the infinite path graph is examined in detail.
As before, without loss of generality, let node $0$ be the sole driver node, and choose nodes $k$ and $\ell$ to be the target nodes.
The output Gramian can thus be written,
\begin{equation}\label{eq:2targ}
  CW(t)C^T = \left[ \begin{array}{cc}
    w_{\ell,\ell}(t) & w_{\ell,k}(t)\\w_{k,\ell}(t) & w_{k,k}(t)
  \end{array} \right]
\end{equation}
where $w_{i,j}$ is defined in Eq. \eqref{eq:chaingram}.
Also, note that because the output controllability Gramian is symmetric, $w_{k,\ell}(t) = w_{\ell,k}(t)$.
The smaller of the two eigenvalues of the matrix in Eq. \eqref{eq:2targ} is,
\begin{equation}\label{eq:lmin}
  \begin{aligned}
  \mu_{\min} &= \frac{w_{\ell,\ell}(t) + w_{k,k}(t)}{2}\\
  &- \frac{1}{2} \sqrt{\left(w_{\ell,\ell}(t) - w_{k,k}(t)\right)^2 + 4 w_{\ell,k}^2(t)}
  \end{aligned}
\end{equation}
\begin{figure}
  \centering
  \includegraphics[width=\columnwidth]{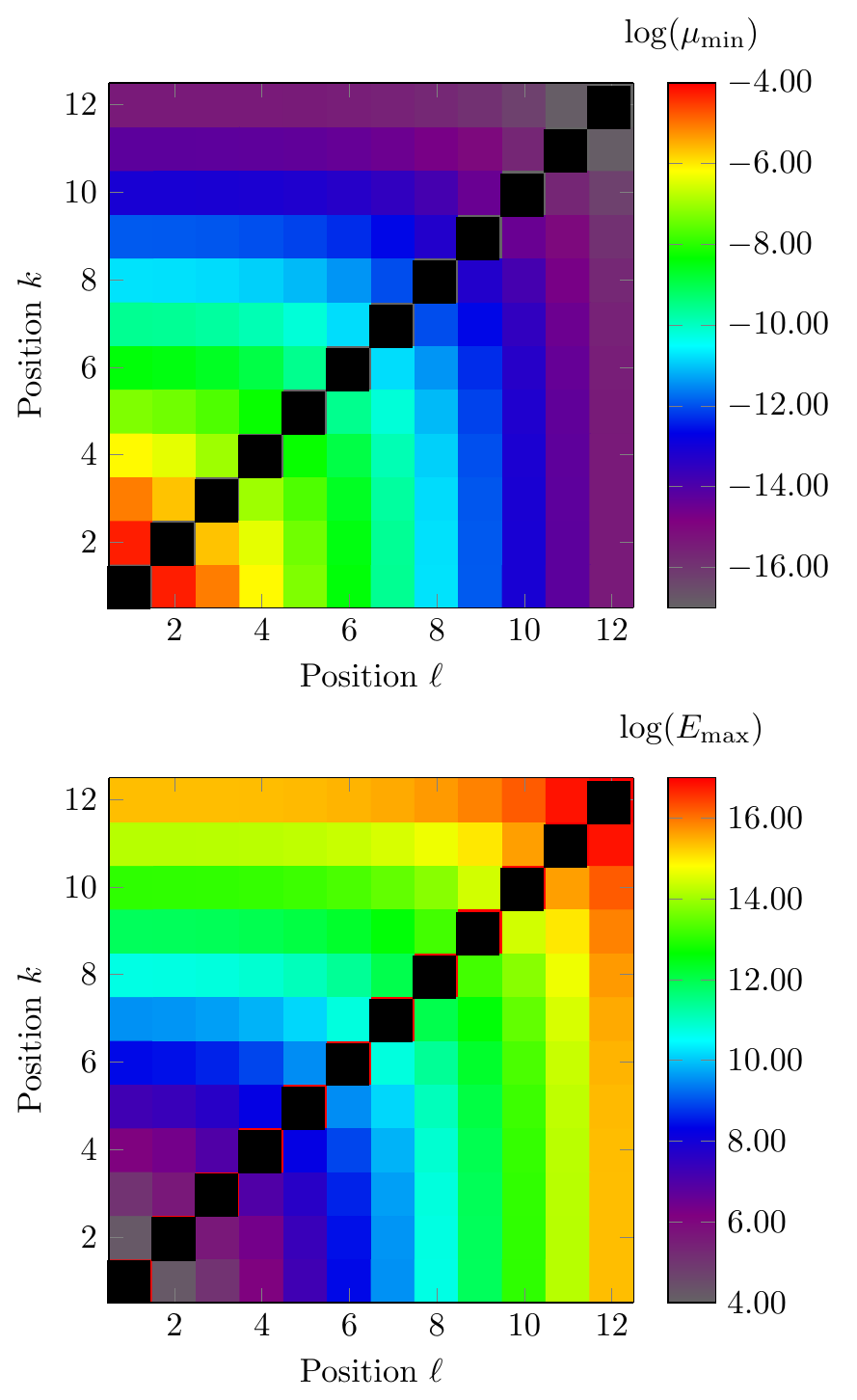}
  \caption{
    Smallest eigenvalue, $\mu_{\min}$, and maximum control energy, $E_{\max} = 1/\mu_{\min}$ for the two target problem on the infinite path graph.
    In both panels, we see that the energy gradient is steeper in the direction corresponding to the target node which is furthest away (horizontal in the lower triangular part and vertical in the upper triangular part.
  }
  \label{fig:two}
\end{figure}
For the case that $w_{\ell,\ell}(t) \gg w_{k,k}(t)$, the value of $\mu_{\min}$ will be approximately the mixed term,
\begin{equation}
  \mu_{\min}(t) \approx w_{\ell,k}(t)
\end{equation}
which behaves more similarly to $w_{k,k}(t)$ then $w_{\ell,\ell}(t)$ as $I_k(z) \ll I_{\ell}(z)$ for $k < \ell$.
This approximation is valid as shown in Fig. \ref{fig:two}(a) where the gradient of the color is much steeper in the direction corresponding to the target node further from the driver node (the horizontal direction in the lower triangular part and the vertical direction in the upper triangular part).
Also shown is the logarithm of the inverse of the smallest eigenvalue in Fig. \ref{fig:two}(b).
The diagonal elements are blacked out as for the case $\ell = k$, the output controllability Gramian becomes singular.
This is expected as when one tries to target two nodes on a bi-directional path graph that are equidistant from the driver node a dilation occurs \cite{yan2015spectrum}, which is equivalent to a singular controllability Gramian.\\
\indent
While previous work \cite{chen2016energy} has been shown numerically that the node most distant from any driver dominates the expression of the control energy, this is the first attempt to explain the cause of such an effect analytically.
\section{Conclusion}
In this paper we derive an analytical expression for the controllability Gramian of the infinite path graph.
We use this result to investigate the scaling of the control energy with respect to the length of the shortest path between a driver node and a target node.
We find that the infinite path result provides an adequate estimate of an upper bound for the control energy for real finite complex networks and that the majority of the energy values for random driver node target node pairs are about an order of magnitude less than the theoretical upper bound.
We conjecture that the lower control energy observed for real networks is due to the redundancy of shortest paths between driver nodes and target nodes.
We have thus considered a \emph{balloon} graph for which we can tune the number of redundant paths between the driver node and the target nodes.
Increasing the number of redundant paths leads to a reduction of the control energy.
Overall we see the balloon graph provides a more accurate prediction of the control energy for the single driver single target problem.\\
\indent
Our results also provide an explanation of the large variance observed when considering the control energy for a particular distribution of target nodes and driver nodes over a network.
It appears the lengths of the paths between the driver nodes and target nodes greatly affect the magnitude of the control energy.\\
\indent
Despite the many substantial results in the literature, there is a current need to develop better analytical tools to characterize the minimum energy control of complex networks.
The framework presented here provides a first step towards an analytical characterization of the control energy in arbitrary networks.
\ifCLASSOPTIONcompsoc
  \section*{Acknowledgments}
\else
  \section*{Acknowledgment}
\fi
This work was funded by the National Science Foundation through NSF grant CMMI- 1400193, NSF grant CRISP- 1541148 and ONR Award No. N00014-16-1-2637.
\bibliographystyle{IEEEtran}

\begin{IEEEbiography}[{\includegraphics[width=1in,height=1.25in,clip,keepaspectratio]{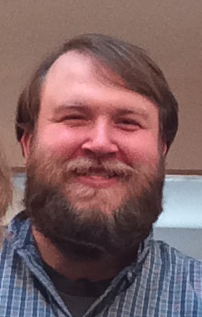}}]{Isaac Klickstein}
received his BS degree in mechanical engineering from the University of New Mexico in 2015.
He is currently a PhD candidate in the Department of Mechanical Engineering at the University of New Mexico.
His supervisor is Dr. Francesco Sorrentino with whom he is conducting research into the optimal control of distributed systems.
His interests include optimal control, numerical methods for optimal control, graph theory, and networked systems.
\end{IEEEbiography}
\begin{IEEEbiography}
[{\includegraphics
[width=1in,height=1.25in,clip,keepaspectratio]{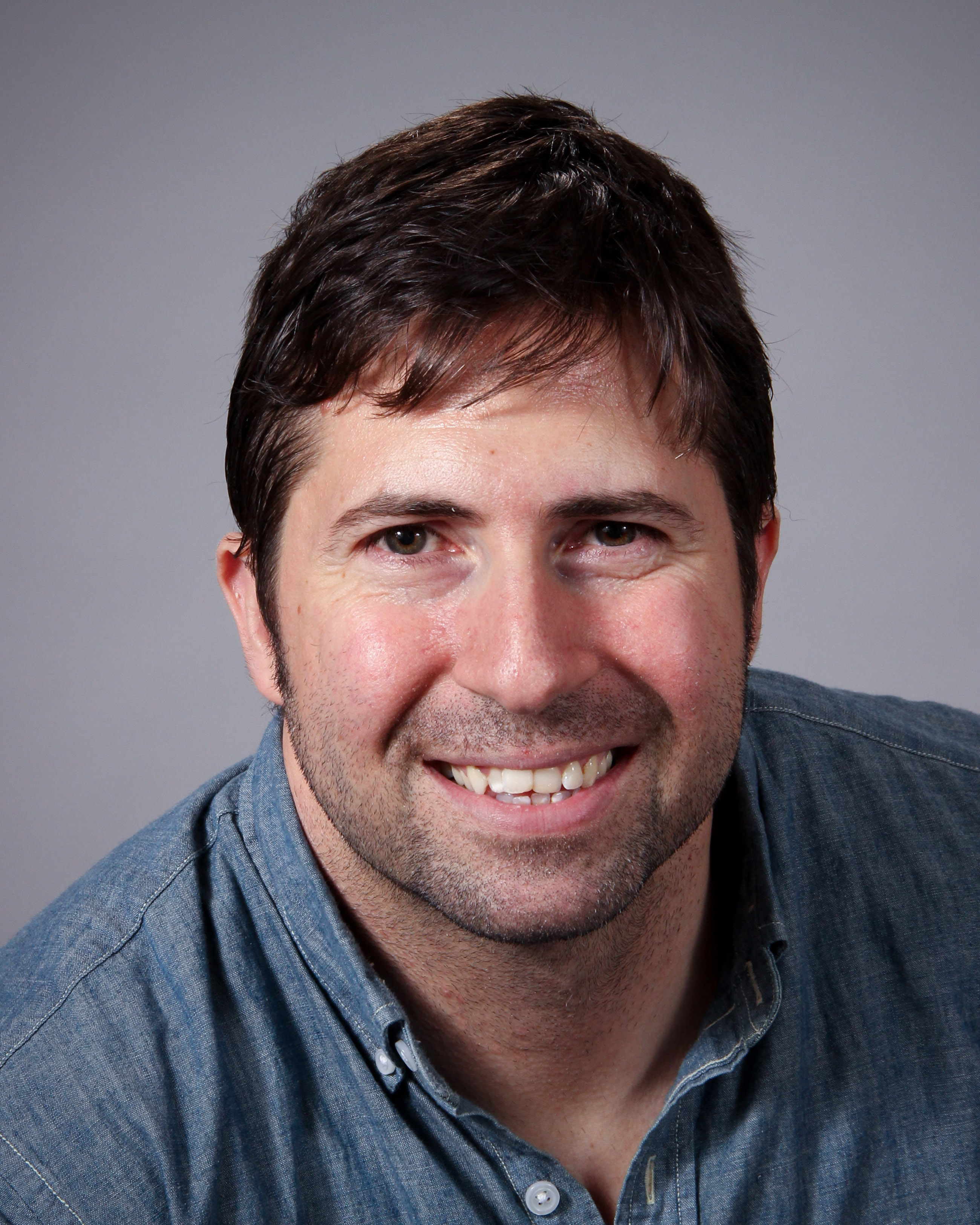}}]
{Francesco Sorrentino}
received a master's degree in Industrial Engineering from the University of Naples Federico II (Italy) in 2003 and a Ph.D. in Control Engineering from the University of Naples Federico II (Italy) in 2007.
His expertise is in dynamical systems and controls, with particular emphasis on nonlinear dynamics and adaptive decentralized control. His work includes studies on dynamics and control of complex dynamical networks and hypernetworks, adaptation in complex systems, sensor adaptive networks, coordinated autonomous vehicles operating in a dynamically changing environment, and identification of nonlinear systems. He is interested in applying the theory of dynamical systems to model, analyze, and control the dynamics of complex distributed energy systems, such as power networks and smart grids. Subjects of current investigation are evolutionary game theory on networks (evolutionary graph theory), the dynamics of large networks of coupled neurons, and the use of adaptive techniques for dynamical identification of communication delays between coupled mobile platforms.
He has published more than 40 papers in International Scientific Peer Reviewed Journals.
\end{IEEEbiography}
\end{document}